\documentclass[11pt]{article}

\usepackage{amsthm,amsmath,amsfonts,amssymb,comment}
\usepackage[authoryear]{natbib}
\usepackage[colorlinks,citecolor=blue,urlcolor=blue]{hyperref}
\usepackage{graphicx}
\usepackage{booktabs}     
\usepackage{subcaption}   
\usepackage{siunitx}      
\usepackage{tabularx}
\usepackage{multirow}
\usepackage{enumerate}
\usepackage{url}
\usepackage[ruled,vlined]{algorithm2e}
\usepackage{color}
\usepackage{tikz}
\usepackage{caption}
\usepackage{changepage}
\usepackage{fancyvrb}
\usepackage{xcolor}
\usepackage{enumitem}

\usetikzlibrary{shapes,decorations,arrows,calc,arrows.meta,fit,positioning}

\theoremstyle{plain}

\theoremstyle{remark}

\newcommand{\probb}{\text{I\kern-0.15em P}}

\title{Multilevel Functional Distributional Models with Applications to Continuous Glucose Monitoring in Diabetes Clinical Trials}

\author{
  Marcos Matabuena\thanks{Universidade de Santiago de Compostela; Department of Biostatistics, Harvard University. Email: \texttt{mmatabuena@hsph.harvard.edu}}
  \and
  Ciprian M. Crainiceanu\thanks{Department of Biostatistics, Johns Hopkins University. Email: \texttt{ccraini1@jhu.edu}}
}

\date{\today}

\begin{document}

\maketitle

\begin{abstract}
Continuous glucose monitoring (CGM) is a minimally invasive technology that measures blood glucose every few minutes for weeks or months at a time. CGM data are often collected in the free-living environment and is strongly related to sleep, physical activity and meal intake. As the timing of these activities varies substantially within- and between-individuals, it is difficult to model CGM trajectories as a function of time of day. Therefore, in practice, CGM trajectories are often reduced to one or two scalar summaries of the thousands of measurements collected for a study participant. To alleviate the potential loss of information, the cumulative distribution function (cdf) of the CGM time series was proposed as an alternative. Here we address the problem of conducting inference on cdfs in clinical trials with long follow up and frequent measurements. Our approach provides three major innovations: (1) modeling the entire cdf and preserving its monotonicity; (2) accounting for the cdfs correlation (because they are measured on the same individual), continuity (results are robust to the choice of the probability grid), and differential error (e.g., medians have lower variability than $0.99$ quantiles); and (3) preserving the family-wise error when the observed data are longitudinal samples of cdfs. We focus on modeling data collected by The Juvenile Diabetes Research Foundation Continuous Glucose Monitoring Group in a large clinical trial that collected CGM data every few minutes for 26 weeks. Our basic observation unit is the distribution of CGM observations in a four--week interval. The resulting data structure is multilevel (because each individual has multiple months of data) and distributional (because the data for each four-week interval is represented as a cdf). The scientific goals are to: (1) identify and quantify the effects of factors that affect glycaemic control in type 1 diabetes patients (T1D); and (2) identify and characterize the patients who respond to treatment.
\end{abstract}


\section{Introduction}\label{sec:introduction}
The continuous improvement and availability of wearable and implantable technology (WIT) provides new scientific and methodological opportunities \citep{liao2019future}. For example, continuous glucose monitors (CGM) \citep{10.2337/diaspect.21.2.112} measure interstitial glucose every few minutes, which can be used for diabetes management. The use of CGM can: (1) help inform  the effect and guide the use of individualized treatments \citep{doi:10.1056/NEJMoa0805017};  (2) be paired with integrated systems of insulin pumps to create artificial pancreas systems \citep{doi:10.1089/dia.2017.0035}; and (3) guide decisions about personalized nutrition due to the individual patient's glycaemic responses to specific food groups \citep{leshem2020gut, ben2021personalized}.

 CGM data are relatively dense (one observation every five minutes) and exhibit substantial within- and between-person variability  and nonstationarity \citep{gaynanova2020modeling}. This is likely due to environmental, biological, and behavioral factors including sleep, physical activity, and meal intake. As the timing of these activities varies substantially within- and between-individuals, it is difficult to model CGM trajectories as a function of time of day. Therefore, in this paper we use the distribution of the CGM observations and not the observed CGM time series. We introduced the term  glucodensities for the probability distribution function (pdf) of CGM data \citep{matabuena2020glucodensities} and here we focus on the cumulative distribution function (cdf). Distributional representation of functional data \citep{Yang2020-of} has been used recently for modeling physical activity based on  accelerometers  \citep{10.1093/jrsssc/qlad007,ghosal2021distributional,ghosal2021scalar, ghosal2023multivariate, ghosal2023predicting, Matabuena2022, koffman2023walking, morris1}, fMRI  \citep{tang2020differences}, and CGM  \citep{matabuena2020glucodensities,cui4investigating}.

We extend these ideas to settings where distributional data are collected at multiple visits for the same participant. 
This structure was considered before in the literature  \citep{cui2021fast,greven2010longitudinal,scheipl2015functional,staicuislam2019}, but not for distributions observed longitudinally. 
For raw CGM trajectories recorded during the night, several authors have proposed multilevel functional models for CGM  \citep{gaynanova2022modeling,sergazinov2022case,matabuena2024multilevelfunctionaldataanalysis}. 
Specifically, \citep{gaynanova2022modeling} introduced a Beta-distribution functional model that captures the non-Gaussian nature of the CGM raw trajectories during sleep. 
More recently, \citep{sergazinov2022case} combined this  strategy with the fast and scalable framework of \citep{cui2022} to address inference on the fixed effects. 
Building on these computational advances, we adapt their massive univariate fitting algorithm to distributional representations. To preserve monotonicity of cdfs we apply a projection idea  inspired by \citep{petersen2021wasserstein} 
to produce increasing cdfs.
Moreover, the methods proposed by 
\citep{gaynanova2022modeling,sergazinov2022case,matabuena2024multilevelfunctionaldataanalysis}
are designed for the observed time series, which may work well  during periods of sleep (as considered in these papers), when physical activity and meal intake have a smaller effect on the glucose response. However, here we focus on the CGM recordings over the course of the entire day for months, and the timing of physical activity, sleep, and meal intake is unknown. Therefore, we propose instead to focus on the CGM distributions, which preserve  critical information on glucose metabolism and ignores/loses the time-of-day information. These distributions contain all the summary metrics of CGM time series data used in practice, such as the mean, median, or time in range.

  	Distributional representations of time series data are typically obtained over a period of time when disease evolution is unlikely and the observed variability can be assigned to natural and reversible biological changes. For example, in studies of diabetes using CGM monitoring, one could reasonably assume that the disease does not progress too much during one month, but detectable differences can be observed over longer periods of monitoring. This partition leads to a longitudinal study where the measurement at every time point (in this case a four week interval) is a distribution (of CGM measurements over a four-week period). This is a new  methodological problem inspired by a new data structure generated by a new and important scientific study. In particular, there is substantial interest in identifying the factors that affect the longitudinal dynamics of glucose and quantifying their effect on glucose control.  To address this challenge, we propose the first multilevel functional model where the response is the cumulative distribution function (or, equivalently, the quantile function), and the predictors are scalar variables.

    The methods are motivated by and applied to a clinical trial conducted by the Juvenile Diabetes Research Foundation Continuous Glucose Monitoring Study Group. The study evaluated the efficacy of CGM in controlling glucose in study participants with type 1 diabetes compared to a group of study participants with type 1 diabetes who were blinded to the CGM information. The scientific problem is to investigate how baseline patient characteristics and CGM monitoring impact glycaemic control. The study duration was 26 weeks, which for this analysis was partitioned into six four-week intervals and a single two-week period ($26=6\times 4 +2$). The quantile function was calculated for every study participant and every interval to reduce the within- and between-day variability of the quantile function estimation.  

Our approach provides three major innovations: (1)   modeling  the entire cdf and preserving its monotonicity; (2) accounting for the cdfs correlation (because they are measured on the same individual), continuity (results are robust to the choice of the probability grid), and differential error (e.g., medians have lower variability than $0.99$ quantiles); and (3) preserving the family-wise error when the observed data are longitudinal samples of cdfs. The goal is to provide a new set of methodological tools for analyzing modern CGM data sets designed to detect subtle alterations in the glucose--metabolism homeostasis.

	\section{CGM data description}
	\label{sec:description}
	
	Continuous Glucose Monitoring (CGM) is thought to be an important component of the future of diabetes management \citep{10.2337/diaspect.21.2.112, 10.2337/dc23-1137}. Several studies 	
	\citep{doi:10.1056/NEJMoa0805017,https://doi.org/10.1111/imj.13770,Schnell2017,doi:10.1089/dia.2017.0035,Ajjan2017,doi:10.1089/dia.2015.0417,doi:10.1089/dia.2014.0378,doi:10.1089/dia.2012.0079,10.2337/dc18-1444,Ajjan2019,Schnell2017,10.2337/diaspect.21.2.112} have provided evidence that the use of CGM in randomized clinical trials is feasible and can help improve glycaemic control. CGM data are often analyzed using summaries \citep{battelino2022continuous} (such as the daily or weekly mean), which can result in substantial loss of information and may lead to clinical decisions that do not account for the complexity of the data. Indeed, it is possible to have two individuals with similar mean CGM, but with very different profiles \citep{yoo2020time,martens2021making}: one stable around the mean and another with long excursions into the hyper- and hypoglycaemic ranges. These two individuals would require different interventions, but their  summarized data would be indistinguishable. Therefore, here we focus on the quantile function over a period of time (e.g., four weeks) which includes information about the mean, standard deviation (variability), and time spent in all CGM ranges.

	Our research was motivated by data collected by The Juvenile Diabetes Research Foundation (JDRF) Continuous Glucose Monitoring Study Group \citep{doi:10.1056/NEJMoa0805017,juvenile2009effect}, one of the first studies to evaluate the potential of continuous glucose monitoring in the management of type $1$ Diabetes Mellitus (T1DM). Data were obtained from a multi-center clinical trial\footnote{JDRF Continuous Glucose Monitoring (JDRF CGM RCT) NCT00406133 \url{https://public.jaeb.org/datasets/diabetes}}. A total of   $451$ adults and children were randomized into two groups. The first group (treatment) received instructions for using and managing their glucose values using CGM information. The second group (control) was provided standard of care, which included home monitoring, but no CGM information. In this paper we consider a subset of $349$ study participants who had enough longitudinal CGM information to construct at least six functional quantile profiles. The study enrolled study participants with baseline glycated hemoglobin (HbA1c) between $5.7$\% and $10$\%. HbA1c is the primary biomarker for the diagnosis and control of diabetes. According to the \href{https://www.cdc.gov/diabetes/managing/managing-blood-sugar/a1c.html}{US Center for Disease Control} ``a normal A1C level is below 5.7\%, a level of 5.7\% to 6.4\% indicates prediabetes, and a level of 6.5\% or more indicates diabetes".

	The number of scheduled contacts with study staff  was identical for all study participants. Visits were 
	conducted at $1$, $4$, $8$, $13$, $19$, and $26$ weeks ($\pm 1$ week), with one telephone contact between each visit, to review glucose data and adjust diabetes management. The primary outcome of the clinical trial   was the change in the  mean HbA1c from baseline to 
	twenty-six weeks, as determined by a central laboratory.
	
	The first two publications based on these data focused on separate analyses for study participants with  (HbA1c) greater than $7$\% ($n_2=322$ study participants) and less than $7$\% ($n_1=129$ study participants), respectively.   Based on the subset of $322$ study participants with HbA1c $\geq 7$\% the authors reported that: (1) the changes in  HbA1c varied substantially by age group (p-value=$0.003$); (2) there was a statistically significant difference in HbA1c change between groups (continuous monitoring group versus control) for individuals who were 25 years or older (mean difference in change, -$0.53$\%; $95$\% confidence interval [CI], -$0.71$ to -$0.35$; p-value$<0.001$); (3) there was not a statistically significant difference between groups for individuals younger than $24$. For the secondary outcome, glycosylated hemoglobin, a statistically significant difference was found between the groups that used and did not use information from continuous monitoring among individuals who were older than $25$. The difference was not statistically significant for individuals younger than $25$. In a subsequent paper, \cite{juvenile2009effect} focused on study participants with baseline HbA1c less than $7$\% and reported a benefit for using CGM information for controlling hypoglycaemia. However, hypoglycaemic control depends on insulin therapy (in this case, pump systems or injections). As a larger proportion of patients use pump-systems in the treatment compared to the control group ($93$\% versus $79$\%), additional analyses may be necessary to understand whether differences are due to treatment (information about CGM) or to higher use of insulin pumps in the treatment group. 
	
	In this paper we analyse all patients and investigate the distribution of CGM instead of HbA1c as outcome of the study. More precisely, we partition the $26$ weeks of follow up into six four-week and one two-week intervals. In each interval we obtain the quantile function of the CGM data in that particular interval, which provides information about the time spent in any glucose range. The quantile of the time series in every interval is treated as a function. We propose to use longitudinal function-on-scalar (FoSR) regression \citep{cui2021fast,reissfosr,shou2015structured,zipunnikov2014longitudinal} to study the association between CGM profiles and scalar covariates including treatment, age, sex, baseline HbA1c and insulin therapy. The model is functional because we use the entire quantile function as the outcome of interest and longitudinal because quantile functions are observed  over seven periods spanning a total of $26$ weeks.

\begin{figure}[ht!]
\centering\includegraphics[width=0.99\textwidth]{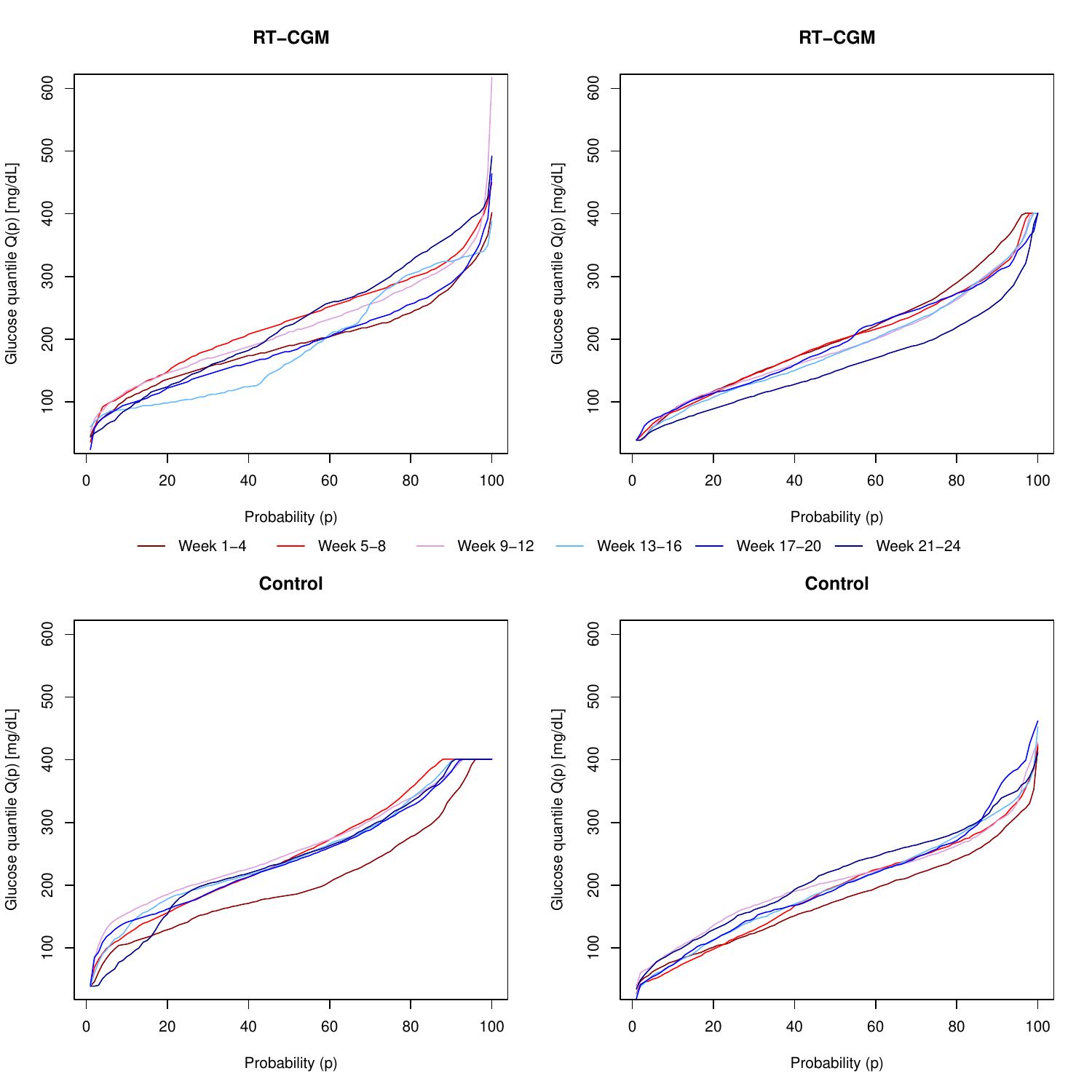}
		\caption{Example of data for four study participants of the Juvenile Diabetes Research Foundation Continuous Glucose  Monitoring Study Group database in their longitudinal evolution of distributional glucose profiles during the first $24$ weeks. Top panels: two patients from the treatment group. Bottom panels: two patients from the control group.}
		\label{fig:grafinitial}
	\end{figure}

	To better understand the structure of the data, Figure \ref{fig:grafinitial} displays the quantile functions for two patients in the treatment (top panels) and two patients in the control (bottom panels) group, respectively. Each one of these participants had only six quantile curves (CGM data for the first $24$ weeks), though other study participants had  seven quantile curves ($26$ weeks). A higher quantile function corresponds to higher blood glucose for longer periods of time. This typically reflects poorer glycaemic control, though in hypoglycaemia ranges (blood glucose below $70$mg/dL) it reflects better glycaemic control. Each quantile function is color coded to provide a temporal representation of the information. The first periods are shown in darker shades of red followed by lighter shades of red, lighter shades of blue, and darker shades of blue, respectively. For example, in the top right panel the quantile functions are higher during the first two periods (week 1-4 and week 5-8) and lower in the two subsequent periods (week 9-12 and week 13-16). The last two periods are different from each other with the fifth period (week 17-20) being higher, though not quite at the level in the first two periods. The sixth period (week 21-24) is much lower than all other periods. Overall, Figure \ref{fig:grafinitial} illustrates that the evolution of CGM quantiles is highly individualized and exhibits substantial heterogeneity within and between individuals.

\begin{figure}[ht!]
		\centering\includegraphics[width=0.99\textwidth]{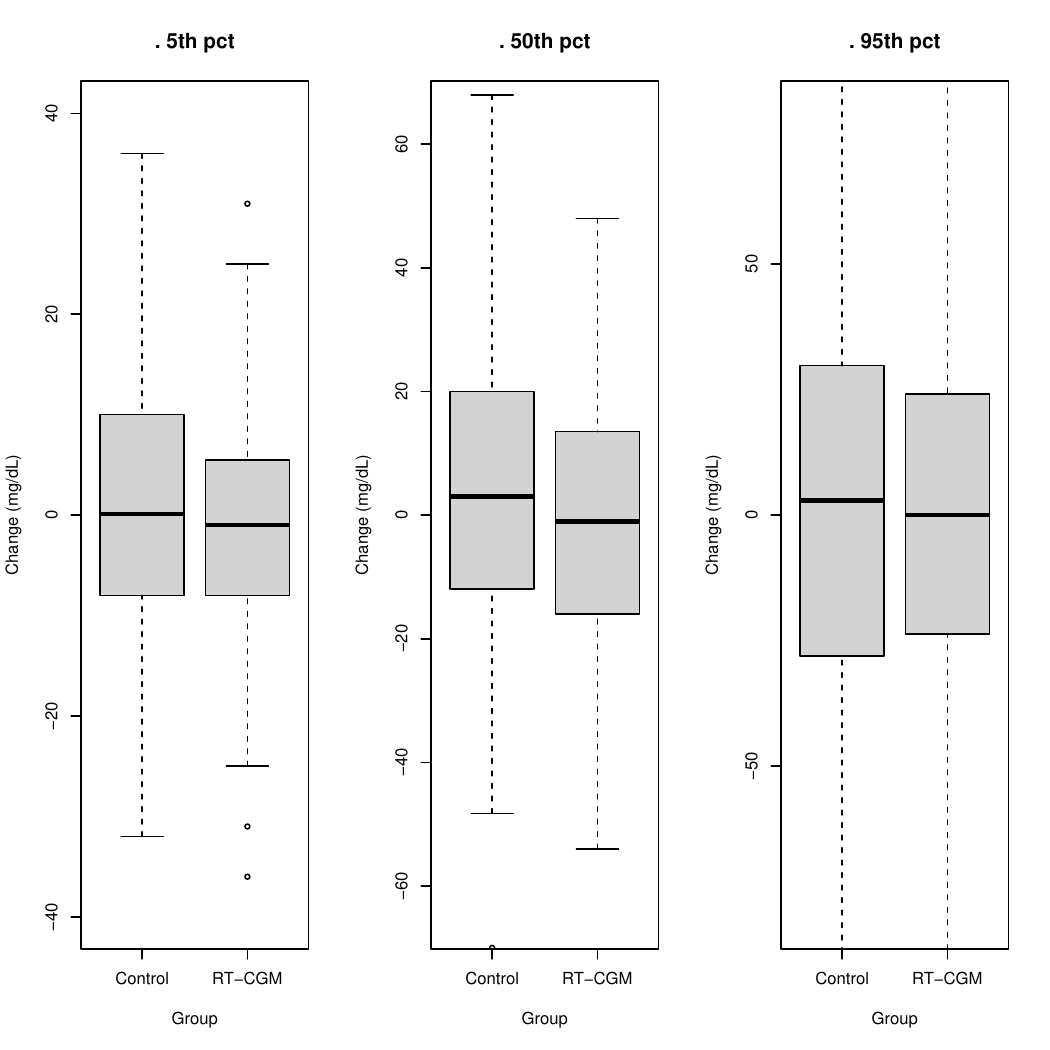}
		\caption{Boxplots of differences (post-pre) in specific quantiles ($0.05$, $0.50$, and $0.95$) by Control and Treatment group (RT-CGM).}
		\label{fig:graf2}
	\end{figure}

As longitudinal functional data are quite difficult to visualize over so many study participants, we display data for a few selected CGM quantiles. More precisely, we focus on the difference between the quantiles corresponding to probabilities $0.05$, $0.50$, and $0.95$ at the end and beginning of the trial. Figure \ref{fig:graf2} displays boxplots of the differences between the last and first period in the corresponding quantiles  in the treatment and control groups, respectively. Therefore, these boxplots contain information only from the first and last periods, and not from the other periods. The goal of this paper is to provide a class of models to analyse all quantiles simultaneously, not just the $0.05$, $0.50$, and $0.95$ quantiles,  while accounting for all longitudinal information and specific covariates.
	
\section{Models for longitudinal distributional representations}\label{sec:long_dist}
	
\begin{figure}[ht!]
		\centering
	\includegraphics[width=0.8\linewidth]{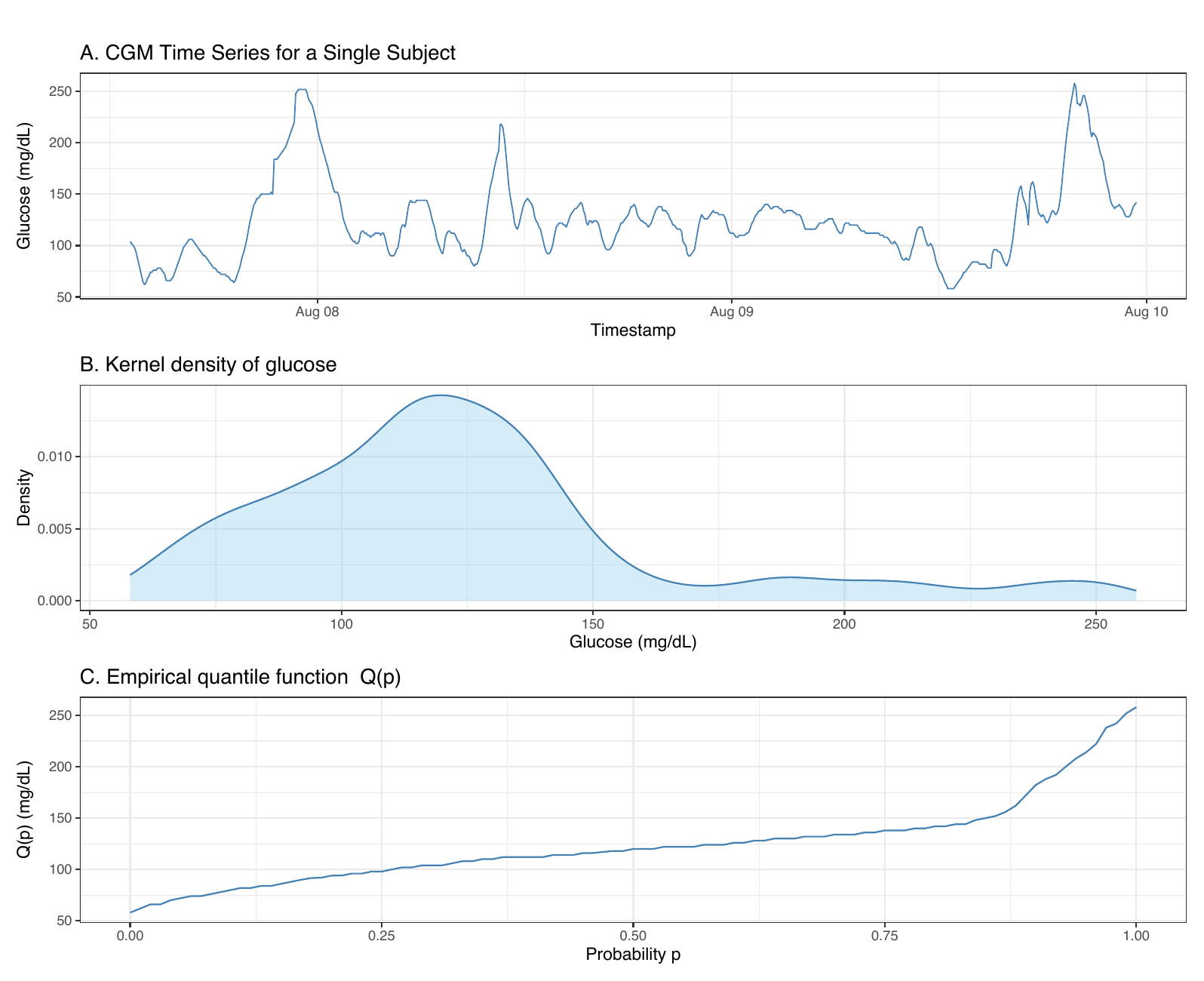}
		\caption{The CGM recording from a normoglycaemic patient, and the corresponding glucodensity, and quantile function.
		}
		\label{fig:gluco_overview}
	\end{figure}

 In this paper we will use the distribution function of CGM data, termed glucodensity \citep{matabuena2021glucodensities,	matabuena2022kernel, matabuena2023hypothesis, cui4investigating}.
	Figure \ref{fig:gluco_overview} provides an example of how the CGM time series (first panel) for a normoglycaemic study participant is transformed into a density function, and the corresponding quantile function. Formally, the space of glucodensities is defined as $\mathcal{D}=\{f\in L^{2}\left(\left[40,600\right]\right): \int_{40}^{600} f\left(s\right) ds=1 \hspace{0.2cm} \text{and} \hspace{0.2cm}  \int_{40}^{600} f^{2} \left(s\right) ds<\infty \}$. The values $40$ and $600$ mg/dL are selected to span the full physiologically-plausible range of glucose values, although they can be adjusted for specific applications.

For a collection of CGM measurements, $\{Y_{ij}\}_{j=1}^{n_i}$, for study participant $i$, our outcome of interest is the empirical quantile function $\widehat{Q}_{i}(p)=\inf\bigl\{g\in[40,600]:\widehat{F}_{i}(g)\ge p\bigr\}$, for $p\in[0,1]$, where
$\widehat{F}_{i}(g)=\frac{1}{n_i}\sum_{j=1}^{n_i}\mathbb{I}\!\bigl\{Y_{ij}\le g\bigr\}$ for  $g\in[40,600]$ is the  corresponding empirical cumulative distribution function.

	\subsection{Longitudinal Functional Quantile Models}\label{sec:multilevel1}
	
	The clinical trial study was partitioned in six four-week periods and one two-week period for a total of seven periods. In every period CGM data are sampled once every five minutes and then transformed into a quantile function; so for every four-week period there are roughly $8{,}000$ observations per person and for every two-week period about $4{,}000$. These data are then summarized by the quantile function resulting in up to seven quantile functions per study participant (as some study participants do not have data in the last two-week period) that are indexed by time. This results in a longitudinal (because periods are indexed by time) distributional (because the measurement is the quantile function over a period) data structure. We are interested in studying the association between study participants' covariates and the CGM quantiles observed longitudinally, while accounting for baseline, longitudinal and white noise variability \citep{10.1093/biostatistics/kxs051}.
	
The structure of the problem is akin to the one introduced by \citep{Greven2010}. However, their primary focus was on the longitudinal decomposition of functional variability without imposing restrictions on the functional data or addressing inference for covariate effects. In the context of functional multilevel data analysis using CGM data, the methods and scientific problems are similar to the ones introduced recently by \citep{https://doi.org/10.1111/biom.13878}. The difference is that we focus on the empirical quantile process, which reduces the effect of exact timing of activity and meal intake on the CGM response. Moreover, we focus on long-term changes in patients over $26$ weeks, which is a period when physiological changes are thought to be possible.

Our functional data consist of quantile functions, which are inherently increasing functions. The main focus of the paper is identifying and quantifying the effects of covariates on CGM quantiles. The problem is  related to the one described by \citep{goldsmith2012,10.1093/biostatistics/kxp058}, though their emphasis was on longitudinal scalar-on-function rather than function-on-scalar regression. To the best of our knowledge, this is the first use of these methods in the context of CGM data for clinical diabetes trials using distributional representations of long-term CGM time series.

	\subsection{Longitudinal function-on-scalar regression models}\label{sec:multilevel_with_cov}

We first introduce some notation. Denote by $Q_{ij}(p)$, $p\in [0,1]$, the random quantile function for individual $i=1,\ldots, n$ during period $j=1,\ldots, J_i$. For notation simplicity we assume that $J_i= J$ for all $i=1,\ldots, n$, though methods can account for a different number of observations per study participant. We also observe time-dependent and time-independent covariates such as demographics and HbA1c. To account for these  covariates we consider models of the type

\begin{equation}
  Q_{ij}(p)= \sum_{l=1}^L X_{ijl}\beta_l(p) + \sum_{k=1}^K Z_{ijk}U_{ik}(p)  +W_{ij}(p)\;, 
  \label{eq_fast}
\end{equation}
\noindent where $X_{ijl}$ are $L$ fixed-effects covariates, $\beta_l(p)$ are the fixed-effect parameter for covariate $X_{ijl}$  probability $p$ quantile, $Z_{ijk}$ are $K$ random-effects covariates, $U_{ik}(p)$ is a random functional effect corresponding to subject $i$ at probability $p$, and $W_{ij}(p)$ is the residual variation at visit $j$ that is unexplained by the subject-specific fixed and random effects. We assume that the $U_{ik}(\cdot)$ and $W_{ij}(\cdot)$ processes are zero-mean square-integrable processes, with $W_{ij}(\cdot)$ being uncorrelated with all $U_{ik}(p)$, though the $U_{ik}(p)$ can be correlated among themselves. We do not impose any additional restrictions on these processes, though the outcomes $Q_{ij}(p)$ are increasing functions of $p$. We propose to fit the models by ignoring this fact, produce predictions of $Q_{ij}(p)$, and then project these predictions onto the space of increasing functions.  

Models such as \eqref{eq_fast} have been proposed in the literature before, are easy to write down, but are difficult to fit in real-life applications. To address this problem, we adapt the recently proposed fast univariate inference (FUI) for longitudinal functional data analysis by \citep{cui2021fast}. This approach can be implemented by fitting many pointwise mixed-effects models and then smoothing the fixed-effects parameters over the functional domain. For practical purposes, we assume that $U_{ik}(\cdot)$ and $W_{ij}(\cdot)$ are Gaussian processes. Consequently, we leverage the capabilities of software designed for Gaussian univariate outcomes, such as the \texttt{lme4} \textbf{R} package. The following algorithm was used to fit model~\eqref{eq_fast}:
	
	\begin{enumerate}
        \item For each point $p\in T_{m}=\{p_{1},\dots,p_{m}\}$, fit a separate point-wise  linear mixed model using standard multilevel software for Gaussian responses, that is
		
		\begin{equation*}
		Q_{ij}(p)= \sum_{l=1}^L X_{ijl}\beta_l(p) + \sum_{k=1}^K Z_{ijk}U_{ik}(p)  +W_{ij}(p).
		\end{equation*}

		\item Smooth the estimated fixed-effects coefficients $\widetilde{\beta}_l(p)$ using a linear smoother $\widehat{\beta}_l(p)= S_l\widetilde{\beta}_l$, where  $S_l$ is a smoother that may or may not depend on $l$.

		\item Use a bootstrap of study participants to conduct model inference:
		
		\begin{enumerate}
			\item Bootstrap the study participants $B$ times with replacement. Calculate $\widehat{\beta}_l^{b}(p)$, the estimator of $\beta_l(p)$ conditional on the $b=1,\ldots,B$ bootstrap sample.

   \item Arrange the $\widehat{\beta}_l^{b}(p)$ estimators in a $B\times m$ matrix (bootstrap samples by probabilities) and obtain the column mean  $\bar{\beta}_{bl}(p)$ and variance $v_{l}(p)={\rm Var}\{\beta_{bl}(p)\}$ estimators.
			
			\item Conduct a Functional Principal Component Analysis (FPCA) on the $B\times m$ dimensional matrix, extract the top $Q$ eigenvalues $\lambda_{1l},\dots \lambda_{Ql}$ and corresponding eigenvectors $\gamma_{1l},\dots \gamma_{Ql}$.
			
			\item For $h=1,\dots, N_{iter}$ do
			
			\begin{itemize}
				\item Simulate independently $\xi_{hq}\sim N(0,\lambda_{ql})$ for $q=1,\dots, Q$. Calculate $\widehat{\beta}_{l,h}(p)= \bar{\beta}_l(p) +\sum_{q=1}^{Q} \xi_{hq} \gamma_{ql}$.
				\item Calculate $u_{hl}= \max_{p\in [0,1]}   \left\{\left|\widehat{\beta}_{l,h}(p)-\bar{\beta}_l(p)\right|\Huge/\sqrt{v_{l}(p)}\right\}$
			\end{itemize}
			
			\item Obtain $q_{1-\alpha,l}$ the $(1-\alpha)$ empirical quantile of the $\left\{u_{1l},\dots,u_{{N}_{iter}l} \right\}$ sample.
			
			\item The joint confidence bands at $p$ are calculated as $\widehat{\beta}_l(p)\pm q_{1-\alpha,l} \sqrt{v_{l}(p)}$.
		\end{enumerate}
	\end{enumerate}

The bootstrap stage resamples subjects to account for the within-person correlation of measurements. After resampling, we apply a smoothing step to the pointwise bootstrap estimates and pointwise as well as correlation and multiplicity adjusted (CMA) confidence bands are obtained as described in \citep{FDAwithR}. The massive univariate fitting methodology has been used before in the context of  density functional responses \citep{petersen2021wasserstein,brito2022analysis, GHOSAL2023107614}. Here we focus on more complex functional mixed effects models to account for the longitudinal sampling of distributions. The first use of this estimation and testing approach for fixed effects in multilevel functional data can be traced to \citep{Crainiceanu2012}. We also use projections on the space of increasing functions to account for the monotonicity of the quantile functions. To the best of our knowledge, this is the first time the problem is formulated and no other methods currently exist for addressing it. Our approach is computationally efficient for models that are notorious for computationally difficulty; for more discussions, see \citep{scheipl2015functional,matabuena2022estimating, cui2021fast}.

A natural question to ask is ``Why does the ``massive--univariate'' fit work?''
First, pointwise estimation yields \emph{consistent} estimators of $\beta_l(p)$
under the assumption of independent subjects, \emph{regardless} of the 
correlation within each curve.
Second, because the model is \emph{linear in the parameters}, the estimating
equations at each $p$ depend on the random effects only through their first two
moments (mean and variance).  Hence, the same pointwise procedure works for
any random-effect structure—random intercepts, random slopes, nested
or crossed effects, serial correlation—\emph{without assuming Gaussianity}.
Finally, any efficiency loss from ignoring intra-curve dependence is mitigated
post-hoc by: (i) smoothing the pointwise estimates; and (ii) constructing
simultaneous confidence bands via a subject-level bootstrap that preserves the functional dependence.

The proposed algorithm relies on three components: (i) a grid of probabilities $T_M=\{p_1,\dots,p_M\}$; (ii) the individual quantile functions $Q_{ij}(\cdot)$ for each subject $i = 1, \dots, n$ and visit $j=1\dots, n_i$; and (iii) a linear smoother $S_\ell$ for every fixed effects function $\beta_l(\cdot)$, $l=1,\ldots,L$. Because the raw glucose trajectories are continuous and smooth, and the monitoring range in our study is equal to $[40, 600]$, an equally spaced grid of $M=100$ points was considered sufficiently dense for practical estimation. Each  empirical quantile function is a consistent estimator of the true quantile function under both $\alpha$- and $\beta$-mixing dependence conditions. The smoother $S_\ell$ was a fixed cubic B-spline basis with $12$ degrees of freedom. However, any reasonable smoother or even not smoothing could be used \citep{cui2022}.

\section{Simulations}\label{sec:simulations}
Simulations were conducted to evaluate the inferential performance of the model described in Section~\ref{sec:multilevel_with_cov} for different scenarios. Each scenario involved $N_s=1000$ simulations. Since this is the first model of longitudinal distributional analysis, there were no other simulation studies to emulate or methods to compare to. The purpose of our simulations was to evaluate the finite-sample performance of the proposed methods in two respects:
(i) the estimation accuracy of the functional fixed-effect coefficients; and
(ii) the inferential performance of the confidence bands.

\subsection{General structure of the simulation study}\label{subsec:general_structure}

We consider two different simulation scenarios: (1) generate data directly from a multilevel quantile process that contains individual and visit effects, but the fixed effects do not depend on a vector of covariates, $X$; and (2) generate data from a quantile process that depends on a vector of covariates $X$. In both scenarios, simulations were conducted for different sample sizes.

For each simulation \(s = 1,\dots,N_s = 1000\) and for every fixed-effect
function \(\beta_l(p)\) (\(l = 1,\dots,L\)), we compute the global and
pointwise mean squared error (MSE):

\[
\widehat{\text{MSE}}_{s,l} =
      \int_0^1 \{\beta_l(p)-\widehat{\beta}_{s,l}(p)\}^2\,dp, 
\qquad
\widehat{\text{MSE}}_{s,l}(p) =
      \{\beta_l(p)-\widehat{\beta}_{s,l}(p)\}^2.
\]

\noindent Here, \(\widehat{\beta}_{s,l}(p)\) denotes the estimator of \(\beta_l(p)\) coefficient
 under model~\eqref{eq_fast} in the \(s\)-th replication. To summarize performance over \(N_s\) runs, we use the median (over $s$) of the integrated MSE (for every $l$)
pointwise MSE (at every $p$ and for each $l$).

To asses the inferential properties of our estimators  we use the coverage frequency of the true functions for both the joint and pointwise confidence band defined as follows:


\[
f^{\alpha,\text{joint}}_{l} = \frac{1}{N_s}\sum_{s=1}^{N_s}\mathbb{I}\left\{\beta_{l}(p) \in \widehat{\mathcal{C}}^{\alpha, \text{joint}}_{s,l}(p),\hspace{0.1cm} \forall p\in [0,1]\right\}\;,
\]

\noindent and

\[
f^{\alpha,\text{pointwise}}_{l} =
\frac{1}{N_s}\sum_{s=1}^{N_s}
\frac{1}{M}\sum_{m=1}^M
\mathbb{I}\left\{\beta_{l}(p_m) \in \widehat{\mathcal{C}}^{\alpha, \text{pointwise}}_{s,l}(p_m)\right\}\;,
\]

\noindent where $\widehat{\mathcal{C}}^{\alpha, \text{joint}}_{s,l}(p)$ and $\widehat{\mathcal{C}}^{\alpha, \text{pointwise}}_{s,l}(p)$ are the CMA (joint) and unadjusted (pointwise) $100(1-\alpha)$\% confidence bands.

\subsection{Direct simulation of the multilevel quantile process without covariates}\label{subsec:direct_sim}
Consider the following data-generating process

 \begin{figure}[ht]
    \centering
     \includegraphics[width=0.95\textwidth]{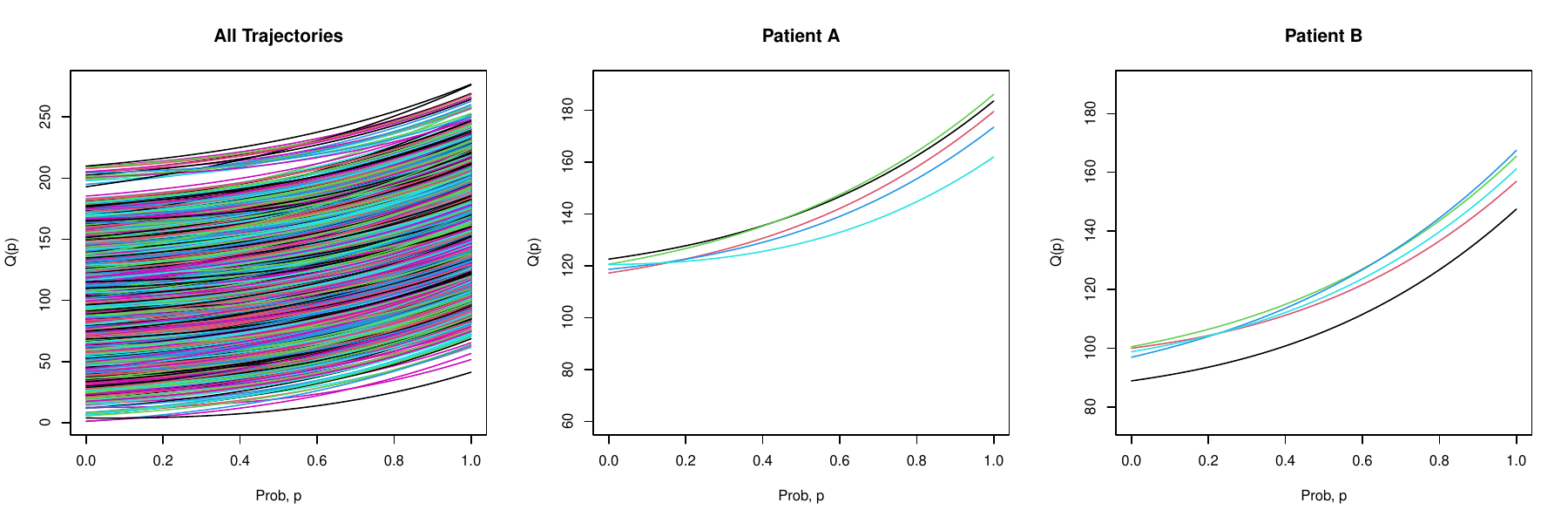}
    \caption{Simulated quantile functions from model~\eqref{eq:direct_simulations} for $n=500$ study participants with a total of $2500$ visits ($J=5$ visits per person) and $\rho = 0.3$. Left panel: all trajectories; Center and right panels: two study participants.}
    \label{fig:examplesiml}
\end{figure}

	\begin{equation}
	Q_{ij}(p)= 100+20(p+p^2+p^3) +U_i+\frac{10pjW_{ij}}{J}+\epsilon_{ij}\;, 
 \label{eq:direct_simulations}
	\end{equation}
\noindent where $i=1,\ldots, n$ indicates subjects and $j=1,\ldots,J$	indicates visits. Denote by $\mathbf{W}_i=(W_{i1},\ldots,W_{iJ})^t$ the vector of visit-specific random effects and by $U_i$  the subject-specific random effects. We generate $U_i$ and $\mathbf{W}_i$ independently with $U_i \sim N(0, 30^{2})$ and $\mathbf{W}_{i}\sim N\left(0 ,\Sigma\right)$, where all the $(r,s)$ entries of $\Sigma$ are equal to $\rho$ for $r\neq s$ and $1$ for $r=s$. For practical purposes, the quantile process $Q_{ij}(p)$ is approximated using an equispaced grid of $M=100$ elements, denoted as $\tau = \{p_{m}\}_{m=1}^{100}$, where $p_{m}=\frac{m}{100}$ is defined as $\frac{m}{100}$, where $m=1,\ldots,M$.

We consider the following scenarios,  $\rho\in \{0,0.3,0.6\}$, $n\in \left \{300,500,1000,2000\right\}$, $J\in \left\{5,10,20\right\}$, and $\epsilon_{ij}\sim N(0,1)$. This is a particular case of model~\eqref{eq_fast} with one fixed effects  function $\beta_1(p)=100+20*(p+p^2+p^3)$, one subject-specific random effect function $U_{i1}(p)=U_i$ that does not depend on $p$, one random effect covariate $Z_{ij1}=1$, and one random visit/person-specific deviation function $W_{ij}(p)=10pjW_{ij}/J$. 
 
 The left panel in Figure~\ref{fig:examplesiml} displays a sample of $2500$ simulated quantile functions from the model~\eqref{eq:direct_simulations} for $J=5$ and $\rho=0.3$. The center and right panels in Figure~\ref{fig:examplesiml} display the specific quantile functions for two study participants. The data generated by this model maintain some of the characteristics of the observed data, including the range of observations, roughly between $40$ and $250$. This explains some of the choices we have made for the fixed-effects function, as well as for the random-effects distribution, though other choices are possible. The function $\beta_{1}(\cdot)$, which represents the mean quantile profile of the generative quantile process, was defined to emulate a plausible pattern ranging from $Q(0)=100\,\text{mg/dL}$ to $Q(1)=160\,\text{mg/dL}$, thus spanning glucose concentrations range from healthy to hyperglycaemic.

 Table \ref{tab:simulations} reports the median of integrated mean square error, $\widehat{MSE}_{s,1}$ $s=1\dots, N_s$, stratified by correlation coefficient ($\rho = 0,0.3,0.6$) and sample size ($n = 300,500,1000,2000$).
Table \ref{tab:coverage0} presents the empirical coverage of the $95$ \% joint and pointwise confidence bands in $N_s = 1000$ replications for the setting $J = 5$, $\rho = 0.3$ and $n = 300,500,1000$. Due to computational constraints, the cluster bootstrap analysis was performed only for $n = 1000$. The main findings are described bellow:

\begin{table}[ht!]
    \centering
    \small
    \begin{tabular}{l *{9}{r}}
        \toprule
        & \multicolumn{3}{c}{$\rho = 0$} & \multicolumn{3}{c}{$\rho = 0.3$} & \multicolumn{3}{c}{$\rho = 0.6$} \\
        \cmidrule(lr){2-4}\cmidrule(lr){5-7}\cmidrule(lr){8-10}
        $n$ & $J=5$ & $J=10$ & $J=20$ & $J=5$ & $J=10$ & $J=20$ & $J=5$ & $J=10$ & $J=20$ \\
        \midrule
        300  & 1.30 & 1.48 & 1.63 & 1.22 & 1.35 & 1.57 & 1.35 & 1.21 & 1.23 \\
        500  & 0.82 & 0.94 & 0.74 & 0.94 & 0.74 & 0.81 & 0.63 & 0.81 & 0.74 \\
        1000 & 0.40 & 0.48 & 0.35 & 0.38 & 0.35 & 0.42 & 0.42 & 0.42 & 0.34 \\
        2000 & 0.32 & 0.16 & 0.25 & 0.22 & 0.26 & 0.19 & 0.25 & 0.23 & 0.21 \\
        \bottomrule
    \end{tabular}
    \caption{Median of $\widehat{MSE}_{s,1}$ $s=1,\ldots,N_s$ for $\beta_1(p)=20(p+p^2+p^3)$ in model~\eqref{eq:direct_simulations}. Results are organized by correlation coefficients ($\rho\in \{0, 0.3, 0.6\}$), sample sizes ($n\in \{300, 500, 1000, 2000\}$), and number of visits ($J\in \{5, 10, 20\}$).}
    \label{tab:simulations}
\end{table}

\begin{enumerate}
    \item As the sample size \(n\) increases, the median of \(\widehat{MSE}_{s,1}\) for \(s = 1,\ldots,N_s\) decreases rapidly, as expected.

    \item Across all combinations of the correlation coefficient, \(\rho\), and number of visits, \(J\), the median of \(\widehat{MSE}_{s,1}\) remains stable. This suggests that the  model performance does not vary substantially  with the correlation between subjects. Moreover, because the algorithm operates pointwise, its computational cost grows only modestly with \(J\); therefore, the gain in efficiency is not offset by excessive computational demands.

    \item The empirical coverage of the joint and pointwise $95$\% confidence bands are close to the nominal level, are slightly conservative for small samples. 
\end{enumerate}

\begin{table}[!ht]
  \centering
  \caption{Bootstrap 95\% coverage for $\beta_1(p)=20(p+p^2+p^3)$ in model~\eqref{eq:direct_simulations} for three sample sizes
           $n\in\{300,500,1000\}$, $J=5$, $\rho=0.3$.}
  \label{tab:coverage0}

  \begin{tabular}{lccc}
    \toprule
    \multirow{2}{*}{\textbf{Type}} &
      \multicolumn{3}{c}{\textbf{Number of subjects}} \\ \cmidrule{2-4}
    & \textbf{300} & \textbf{500} & \textbf{1000} \\ \midrule
    Joint coverage      & 0.97 & 0.96 & 0.95 \\
    Point-wise coverage & 0.97 & 0.97 & 0.96 \\   
    \bottomrule
  \end{tabular}
\end{table}


\subsection{Direct simulation of the multilevel quantile process with covariates}
\label{subsec:direct_sim}

Consider the following data generating mechanism, where the quantile functions depend on sex (binary variable) and age (continuous):

\begin{figure}[t]
  \centering
  \includegraphics[width=\textwidth]{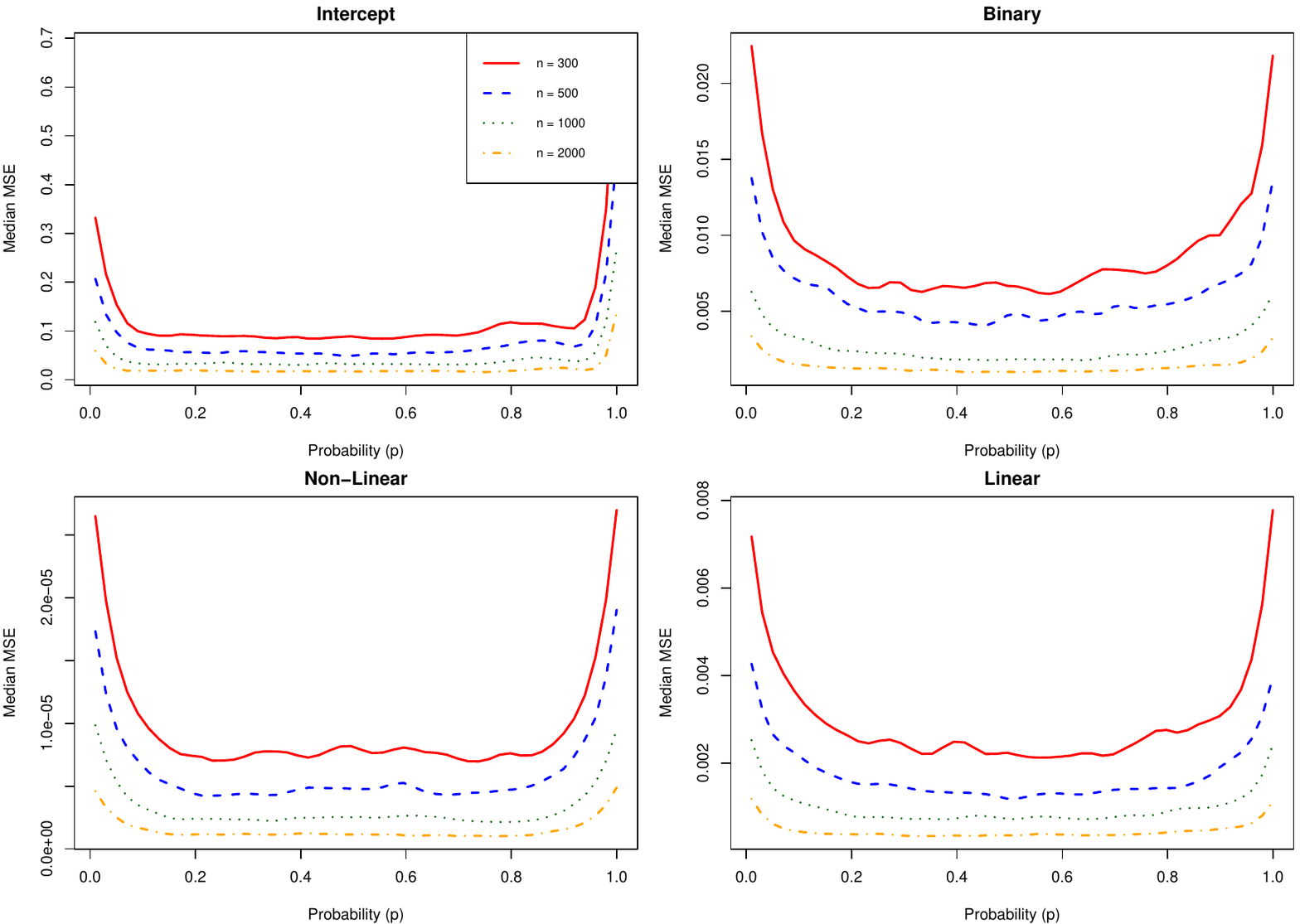}
  \caption{Median pointwise MSE, $\widehat{\mu}(p)$, across
           $N_{s}=1000$ simulations for each coefficient function.
           Columns correspond to sample sizes
           $n\in\{300,500,1000,2000\}$; $J=10$, $L=5$.}
  \label{fig:combined-errors2}
\end{figure}

\begin{equation}
\begin{aligned}
  Q_{ij}(p) &= 100 + 20\bigl(p + p^{2} + p^{3}\bigr)
               + 3  p  \!\Bigl(\sum_{l=1}^{L}X_{il}\Bigr)
               + 20  \,\text{Sex}_{i} \\[4pt]
            &\quad
               + \text{Age}_{i} \Bigl[0.5 + \frac{p}{10}
                   - \Bigl(\frac{p}{5}\Bigr)^{2}\Bigr]
               + U_{i}
               + \frac{10pjW_{ij}}{J}
               + \varepsilon_{ij},
\end{aligned}
\label{eq:quantile_process}
\end{equation}
\noindent where
$U_{i}\sim N(0,3^{2})$,
$W_{ij}\sim N(0,1)$ and
$\varepsilon_{ij}\sim N(0,1)$ are mutually independent.
The covariates are generated independently as
\(
  X_{il}\sim\mathcal{U}(0,3),\;
  \text{Sex}_{i}\sim\operatorname{Bernoulli}(0.55),\;
  \text{Age}_{i}\sim\mathcal{U}(20,80).\)
) The parameter function, $\beta_{\rm sex}(p)=20$, for sex is constant, while the parameter function, $\beta_{\rm age}(p)=0.5+p/10-(p/5)^2$, for age is nonlinear.

We fix $J = 10$, $L = 5$, and assume $\varepsilon_{ij} \sim N(0, 1)$, while varying the sample size $n \in {300, 500, 1000, 2000}$. For each setting, we performed $N_s = 1000$ simulations and computed the median of pointwise mean squared error, \(\widehat{MSE}_{s,l}(p)\) for \(s = 1,\ldots,N_s\) and $p\in [0,1]$. Figure~\ref{fig:combined-errors2} displays the median of the pointwise mean square error  by predictor type (intercept, binary variable, linear effect, nonlinear effect). Table~\ref{tab:coverage0} presents the empirical coverage of the joint and pointwise 95\% confidence bands. Key findings include:


\begin{enumerate}
  \item Across the four types of predictors, increasing the sample size \(n\) results in decreased median pointwise mean square error \(\widehat{MSE}_{s,l}(p)\), as expected.

  \item For moderate sample sizes \(n \in \{300, 500\}\), the error is larger for more extreme quantiles (\(p < 0.1\) and \(p > 0.9\)), which is likely due  to the larger variability of the extreme quantile estimators; this effect tends to be less obvious for \(n = 2000\), as expected.

  \item The pointwise and joint bootstrap confidence bands have close to the nominal level 95\% (Table~\ref{tab:coverage}). For smaller sample sizes (\(n \in \{300, 500\}\)), the coverage rates are slightly higher than nominal.
\end{enumerate}

\begin{table}[t]
  \centering\small
  \caption{Empirical $95\%$ coverage of bootstrap confidence bands
           for the linear, binary, and nonlinear predictors
           ($J=10$, $L=5$).  The first line in each panel is
           \emph{joint} coverage; the second line is
           \emph{pointwise} coverage.}
  \label{tab:coverage}
  \begin{subtable}[t]{0.3\linewidth}
    \centering
    \caption{Linear predictor}
    \begin{tabular}{lccc}
      \toprule
      & 300 & 500 & 1000 \\ \midrule
      Joint      & 0.97 & 0.97 & 0.95 \\
      Pointwise  & 0.96 & 0.97 & 0.96 \\ \bottomrule
    \end{tabular}
  \end{subtable}\hfill
  \begin{subtable}[t]{0.3\linewidth}
    \centering
    \caption{Binary predictor}
    \begin{tabular}{ccc}
      \toprule
      300 & 500 & 1000 \\ \midrule
      0.97 & 0.96 & 0.95 \\    
      0.98 & 0.96 & 0.95 \\    
      \bottomrule
    \end{tabular}
  \end{subtable}\hfill
  \begin{subtable}[t]{0.3\linewidth}
    \centering
    \caption{Nonlinear effect}
    \begin{tabular}{ccc}
      \toprule
      300 & 500 & 1000 \\ \midrule
      0.96 & 0.96 & 0.96 \\    
      0.97 & 0.95 & 0.95 \\    
      \bottomrule
    \end{tabular}
  \end{subtable}
\end{table}

	\section{Results for the Juvenile Diabetes Research Foundation Continuous Glucose  Monitoring Study Group}   \label{sec_results}

        In this section we focus on  the data from the Juvenile Diabetes Research Foundation Continuous Glucose Monitoring Study Group, which was introduced in Section \ref{sec:description}. In particular, we will highlight: (1) that the implementation of a joint analysis of all quantiles is feasible for these data; (2) that this analysis provides different results from a mean or median only analysis; and (3) that building $\alpha$-level confidence bands is possible, intuitive, and easy to implement. To achieve this, in Section~\ref{subsec:scalar_average} we provide a traditional analysis of the average glucose data, while in Section~\ref{subsec:distributional} we provide the analysis based on our multilevel distributional model. This allows a direct comparison of the methods and highlights how our new methods extract more information than standard approaches.

	\subsection{Scalar multilevel functional model}\label{subsec:scalar_average}
	
Let $G_{ij}$ the mean glucose for participant $i$ in the period $j$, with $i = 1,\ldots,n$ and $j = 1,\ldots,J_i$ and the standard linear mixed-effects model:
\begin{equation}
  G_{ij} \;=\; \sum_{l=1}^{L} X_{ijl}\,\gamma_{l} \;+\; b_{0i} \;+\; b_{1i}\,Z_{ij} \;+\; \varepsilon_{ij}\;.
  \label{eq:LMM}
\end{equation}
\noindent Here $X_{ijl}$ ($l = 1,\ldots,L$) are the fixed-effect covariates with corresponding coefficients $\gamma_{l}$, $b_{0i}$ is the participant-specific random intercept; $Z_{ij}$ is time, $b_{1i}$ is the associated random slope, and $\varepsilon_{ij}$ are the residuals. We use the standard assumptions $(b_{0i}, b_{1i})^{\mathsf T} \sim  N(\mathbf 0,\Sigma_{b})$ and $\varepsilon_{ij} \sim  N(0,\sigma^{2})$, and $\varepsilon_{ij}$ are independent of $(b_{0i}, b_{1i})$.

We used $L=11$ fixed effects, including an intercept and ten covariates: RT-CGM (a binary variable indicating whether or not an the individual was randomized to the treatment group), sex (male/female), Weight, Age, Pump (yes/no), period of the year (with one reference category and four factor variables), and HbA1c. Each fixed-effect coefficient $\gamma_l$ represents the expected change in the conditional mean of glucose, $\mathbb{E}\bigl[G_{ij}\mid X_{ij},\,Z_{ij},\,b_{0i},\,b_{1i}\bigr]$, per one-unit increase in $X_{ijl}$, holding all other covariates and the random effects constant.

	\begin{table}[h!]
    \centering
    \renewcommand{\arraystretch}{1.5}
    \begin{tabular}{|l|p{5cm}|}
        \hline 
        \textbf{Variable Name} & \textbf{Description} \\
        \hline
        Age & Age in years of the participant at the time of screening. \\
        \hline
        Group & Binary variable indicating if the patient belongs to the control or treatment group (RT-CGM). \\
        \hline
        Weight & Weight (kg). \\
        \hline
        Gender & Binary variable indicating the gender of the patient. \\
        \hline
        HbA1c & Baseline glycosylated hemoglobin variable. \\
        \hline
        Pump & Binary variable indicating if the patient follows insulin therapy with insulin pump or injections. \\
        \hline
        Seasonal & 
        \begin{itemize}[left=0pt]
            \item Level 1: First seven weeks of the year.
            \item Level 2: Weeks 7-17.
            \item Level 3: Weeks 18-33.
            \item Level 4: Weeks 33-39.
            \item Level 5: Rest of the year.
        \end{itemize} \\
        \hline
        ID & Identification number of the patient. \\
        \hline
        Visit & Visit number of the patient. \\
        \hline
    \end{tabular}
    \caption{Description of the variables used in the Juvenile Diabetes Research Foundation Continuous Glucose Monitoring Study Group.}
    \label{tab:variables_in_analysis}
\end{table}
	
	
\begin{table}[ht!]
\centering
\small              
\begin{tabular}{l r c}
\toprule
\textbf{Variable} & \textbf{Est.\ (95\% CI)} & \textbf{$p$‐value}  \\
\midrule
Intercept   &  80.18 (60.87--99.70)   & $<0.001$  \\
RT--CGM     &  -5.47 (-9.06--\,–1.88) & 0.03      \\
Male        &  -2.60 (-6.17--\, 0.96) & 0.15      \\
Weight      &  -0.06 (-0.17--\, 0.04) & 0.23      \\
Age         &  -0.46 (-0.59--\,–0.34) & $<0.001$ \\
Pump        &  -7.96 (-12.61--\,–3.32) & $<0.001$  \\
Season\_2   &  -1.38 (-3.43--\, 0.66) & 0.18      \\
Season\_3   &  -0.64 (-2.63--\, 1.35) & 0.52      \\
Season\_4   &   0.04 (-2.28--\, 2.36) & 0.97    \\
Season\_5   &   0.10 (-1.71--\, 1.92) & 0.90      \\
HbA1c       &  14.70 (12.50--16.93)   & $<0.001$  \\
\bottomrule
\end{tabular}
\caption{$\beta$-coefficients for fixed numerical and categorical predictors in the conditional mean model of CGM glucose. Conditional $R^{2}=0.72$; marginal $R^{2}=0.35$.}
\label{tab:variables_in_analysis2}
\end{table}

Table~\ref{tab:variables_in_analysis} summarises the covariates included in
model~\eqref{eq:LMM}, while Table~\ref{tab:variables_in_analysis2} reports the
fixed-effect estimates, the $95$\% confidence intervals, and the associated p-values for testing the null hypothesis of no asspciatoon.  Treatment (RT-CGM use), Age, Insulin-pump therapy, and
HbA\textsubscript{1c} have statistically significant associations with the mean
glucose. RT-CGM intervention is associated with a \(-5.47\;\text{mg/dL}\) reduction in average CGM. To better understand the magnitude of this effect, a one unit decrease in HbA\textsubscript{1c} corresponds to a
\(-14.70\;\text{mg/dL}\) decrease in average CGM. This is a decrease roughly three times larger than of the intervention, but should be kept in perspective. Recall that measurements of HbA1c vary in a narrow interval ($5.7$\% to $10$\%), no -diabetes is defined as ${\rm HbA1c}<5.7$\%, while diabetes is defined as ${\rm HbA1c}\geq 6.5$\%. Therefore the difference between diabetes and no-diabetes is $0.9$ HbA1c units, indicating that one unit of HbA1c is a very large clinical change.  A 10-year increase in age was associated with a \(4.60\;\text{mg/dL}\) decrease in average CGM, which indicates that older individuals in this study had better hyperglycaemic control.  Pump therapy is associated with a 
\(-7.96\;\text{mg/dL}\) reduction in average CGM, or roughly $50$\% higher effect than RT-CGM.   The other covariates were not statistically significant at the level $\alpha=0.05$.  The marginal $R^{2}$  was $0.35$ (proportion of variance explained by fixed effects) while the conditional $R^{2}$ was $0.72$ (proportion of variance explained by the fixed and random effects).

\subsection{Distributional Multilevel Functional model}\label{subsec:distributional}
In this section we fit a model with a similar structure with the one in Section~\ref{subsec:scalar_average}, but where the average CGM $G_{ij}$ is replaced by the quantile function $Q_{ij}(p)$ for $p\in[0,1]$. To keep things consistent, we fit the distributional functional model~\eqref{eq_fast} with the same $L=11$ fixed effects used in Section~\ref{subsec:scalar_average}. For each variable \(l\), the functional parameter \(\beta_l(p)\) is modeled as a smooth basis of cubic  B-splines with 12 degrees of freedom in the domain \(p\in[0,1]\), allowing the effect of each covariate to depend on the probability $p$ associated with the quantile $Q_{ij}(p)$. The model also includes a random intercept and slope, with $K=2$, $Z_{ij1}=1$, and $Z_{ij2}=j$ for every $i=1,\ldots,I$ and $j=1,\ldots,J_i$. The functional random intercept and slope, $U_{i1}(p)$ and $U_{i2}(p)$, are allowed to be correlated across probabilities $p$, but are assumed to be independent across subjects, $i$. The subject/period-specific deviations $W_{ij}(p)$ are assumed to be independent across $i$ and $j$ and mutually independent of $U_{i1}(p)$ and $U_{i2}(p)$, respectively. The random effects structure is similar to the LFPCA structure in \citep{Greven2010}. While  \cite{Greven2010} focused primarily on the random effects part of the function, our primary interest is on estimating the smooth fixed effects, while accounting for the complex LFPCA structure of residuals.

	\begin{figure}[ht!]
		\centering\includegraphics[width=0.99\textwidth]{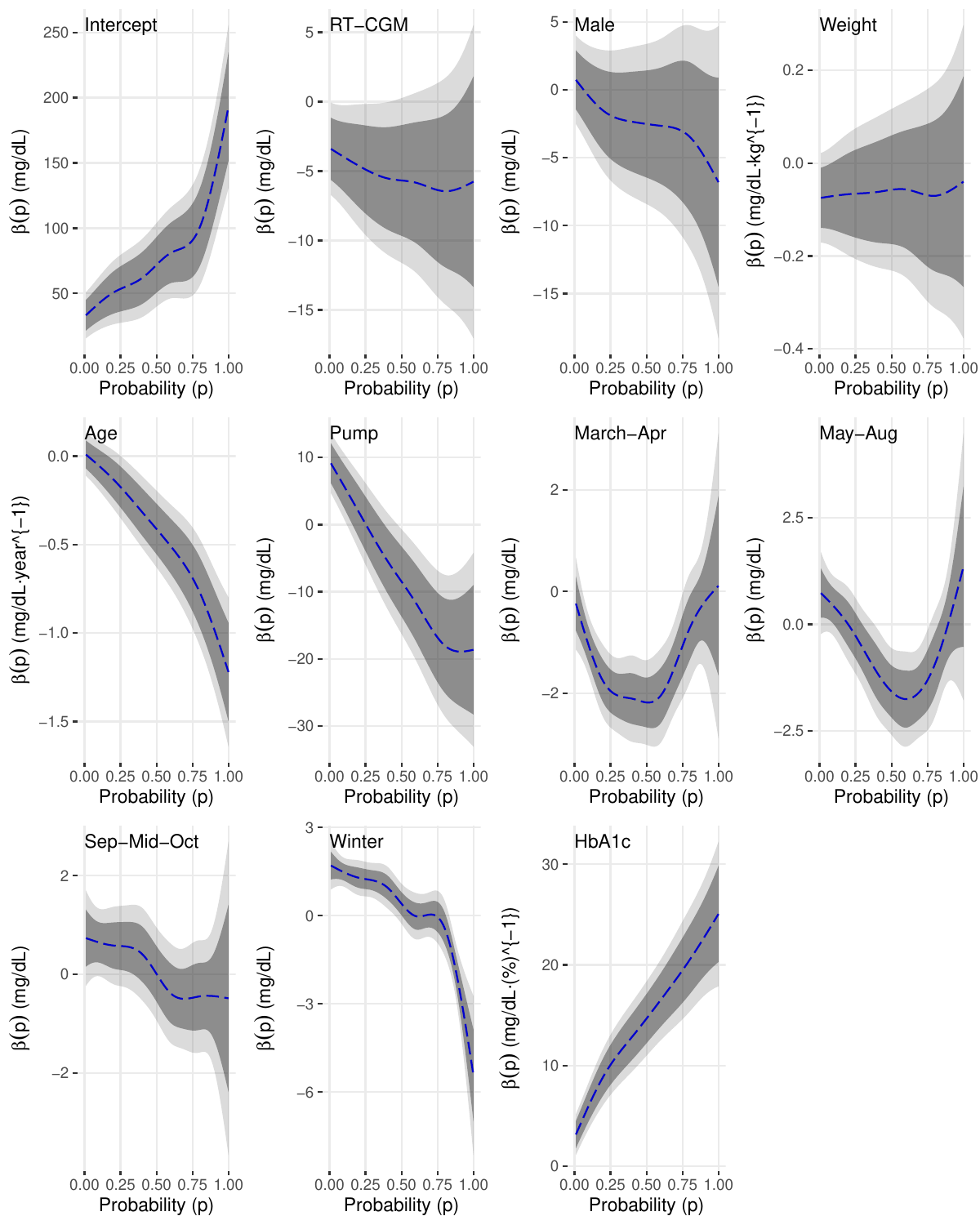}
		\caption{$\beta-$ functional coefficients for the fixed-terms of numerical and categorical predictors in the estimation of the conditional mean for the distributional quantile representation}
		\label{fig:graf3}
	\end{figure}

	Model~\eqref{eq_fast} was fit using the FUI approach described in Section \ref{sec:multilevel_with_cov}. Figure \ref{fig:graf3} displays the estimated functional coefficients (blue dashed curves), $95$\% pointwise confidence bands (dark shade of gray), and the $95$\% correlation and multiplicity adjusted (CMA) confidence  bands (lighter shade of gray) \citep{FDAwithR}.  Recall that smaller probabilities  (lower quantiles) correspond to  the hypoglycaemia range and larger probabilities (upper quantiles) correspond to  the hyperglycaemic range. 
	
	The effect of the RT-CGM variable (treatment group) is shown in the first row, second column panel. The point estimator is consistently negative, indicating a reduction in blood glucose levels in the treatment group across all quantiles. The estimated reduction is more pronounced in the hyperglycaemic ranges, though it is statistically significant (as defined by the 95\% CMA confidence bands) only in the  hypoglycaemic range  of the quantile function ($p<0.4$). In contrast, the scalar model described in Section~\ref{subsec:scalar_average} does not capture that: (1) the effect is mainly due to lowering of the low quantiles of CGM; and (2) the variation in the treatment effect in the higher quantiles is much larger. This is likely due to the much larger variability of the upper quantiles. From a biological standpoint, this may suggest that the RT-CGM intervention may be more effective at curbing extreme glucose excursions, especially in the hypoglycaemic range than at moderating fluctuations around the mean. This finding is novel and was obtained using novel methodology developed in this paper.

    The third and fourth panels in the first row of Figure~\ref{fig:graf3},  correspond to the effects of sex and weight on blood glucose levels for all quantiles. The results indicate that the effects of these variables are not statistically significant after adjusting for the covariates in the study (light shaded areas contain zero for every value of $p$). Similar patterns were observed in the scalar analysis presented in the previous section; here, we demonstrate that these variables are not statistically significant at any level $p \in [0,1]$.

    In the second row, the first panel corresponds to age, indicating a highly significant and neagtive effect of age above $p=0.25$, indicating that older individuals in the trial have better glycaemic control. This effect is roughly linear ass a function of the probability, $p$, indicating much stronger effects closer to $p=1$. For example, an increase of $10$ years in age corresponds to a reduction of $12.5$ mg/dL at the $95$th percentile, whereas the model focusing solely on the mean glucose values indicates an average reduction of $4.6$ mg/dL. This comparison emphasizes that functional quantile analysis provides a deeper, or non-redundant, information to standard models. In our sample patients were between 18 to 35 years, and results suggest that older age in this range is associated with improved glycaemic management, particularly through attenuation of high--glucose excursions that elevate the risk of diabetes complications.

    The second panel in the second row corresponds to the effect of using an insulin pump, and results indicate that the use of a pump corresponds to higher glucose levels in the hypoglycaemic range ($p<0.2$) and lower glucose in the hyperglycaemic range ($p>0.5$). This result is, to our knowledge, novel and provides a detailed fine-mapping of the effects of an insulin pump. This effect could not have been measured by focusing only on the mean CGM. The third and fourth panels in the second row, as well as the first and second panels in the third row, display results for different times of the year relative to the reference period comprising the first seven weeks of the year. These results indicate strong, statistically significant effects of various periods of the year on blood glucose levels. If confirmed, these differences could provide valuable insights into the natural annual variations in glycaemic control that should be considered in clinical decision-making and research applications.
	
	The third panel in the third row of Figure~\ref{fig:graf3} corresponds to the association between baseline HbA1c and measured CGM. The results indicate a very strong and statistically significant association across all quantile levels, as well as a linear increase in the association between probability and CGM levels. The lowest effect was found in the hypoglycaemic range, where an increase of $1$ unit in HbA1c was estimated to be associated with an average increase of $7$ mg/dL in the lowest quantile ($p=0.01$) and an average increase of $20$ mg/dL in the highest quantile ($p=0.99$). While not directly comparable, the mean model provides an estimate of $14.7$mg/dL increase in  average CGM  for a one unit increase in HbA1c. This analysis is also in line with an analysis of median CGM, which would provide an estimate of $15.0$ mg/dL. These results suggest that the association between HbA1c and daily trajectories of CGM is more complex and not completely captured by taking the mean, or median, CGM over the day.
	
	Our analysis complements the original findings \cite{doi:10.1056/NEJMoa0805017,juvenile2009effect} by: (1) providing results across the entire blood glucose range while accounting for covariates and the longitudinal structure of the data; (2) accounting for variables such as insulin pump use that were not balanced by random treatment assignment; and (3) quantifying the seasonal variations of blood glucose for the first time. In particular, our implementation of the new functional distributional method provided evidence that age plays a pivotal role in glycaemic control across all quantile levels. Our findings also demonstrate the effectiveness of Continuous Glucose Monitoring (CGM) in enhancing glycaemic control, particularly for the lower quantiles.
We have identified a linear association between HbA1c levels and the quantile function evaluated throughout its entire domain (i.e., all probability levels), a novel clinical finding in the CGM literature. Most of the literature focuses on the connection between HbA1c and CGM mean and variance but not with the joint CGM quantile functions. While some work connecting HbA1c and time in range exists, this work depends on the choice of the range and does not account for all ranges simultaneously.
	
    From a methodological standpoint, our novel longitudinal distributional/quantile model offers significant advantages over scalar longitudinal analyses because it accounts for the entire distribution. We also introduce an inferential methodology that is relatively easy to understand, apply, and explain because it is closely related to conducting separate analyses of all quantiles. Another important novelty is that we show how to conduct simultaneous inference for all quantiles, while accounting for the complex inherent variability of the distributional processes. The assumptions are also minimal on the quantile processes; for example, we do not require Gaussianity, which would be highly suspect for our data. The new method provides a comprehensive and insightful description of the longitudinal evolution of distributional characteristics of CGM time series over a four-week period, while accounting for patient characteristics.

	\section{Discussion}\label{sec:discussion}

This paper introduces a novel approach to analyse CGM data within the context of clinical trials. Our method leverages distributional representations of CGM data and accounts for the within-person correlation of quantile functions using a  multilevel functional regression model for distributions. We expect that this data type  will become increasingly common in practice and there will be increased need for methods described in this paper. In this paper we focused on the quantile function of glucose profiles derived from Continuous Glucose Monitoring (CGM) technology. However, similar data structures and solutions will appear in various other scenarios where there is a need to summarize high-dimensional functional data as distributions and account for the known,  multilevel or longitudinal structure of the resulting distributions.

Our methods are complementary to the multilevel functional methodology proposed by \citep{gaynanova2022modeling,https://doi.org/10.1111/biom.13878,matabuena2024multilevel}. Indeed, these methods focused on analysing  glucose fluctuations from sleep onset, a period of the day that is not subject to food intake and physical activity. Therefore, these methods cannot be directly extended to CGM during the day because of the heterogeneity of the various events that influence CGM measurements. For these reasons we replaced the CGM time series over multiple weeks with the distribution of the CGM measurements over that period. The downside is that we lose the time-dependence within the four week period, but we gain in interpretability of the CGM measurements during the entire duration of the day.

	\bibliographystyle{apalike}
		\bibliography{article}

\end{document}